\documentclass[twocolumn,nofootinbib,amsmath,amssymb,aps,prd]{revtex4-1}

\usepackage{graphicx}
\usepackage[caption=false]{subfig}

\usepackage{amsmath,amssymb}
\usepackage{amsfonts,amssymb,mathrsfs}

\usepackage[bookmarks=false,pdfstartview=FitH]{hyperref}
\usepackage[all]{hypcap}

\def\be{\begin{equation}}
\def\ee{\end{equation}}
\def\nn{\nonumber}
\def\f{\frac}

\def\pl{{\rm Pl}}
\def\lp{\ell_\pl}
\def\b{\bar}

\def\h{\hat}
\def\t{\tilde}

\def\bra{\langle}
\def\ket{\rangle}
\def\dd{{\rm d}}

\def\de{\delta}

\def\la{\lambda}

\def\ve{\varepsilon}
\def\vp{\varphi}

\def\mH{\mathcal{H}}
\def\mO{\mathcal{O}}

\def\mV{\mathcal{V}}
\def\mI{\mathcal{I}}
\def\mK{\mathcal{K}}
\def\mL{\mathcal{L}}

\usepackage{color}

\begin{document}

\pagestyle{plain}

\title{A relational Hamiltonian for group field theory}

\author{Edward Wilson-Ewing} \email{edward.wilson-ewing@unb.ca}
\affiliation{Department of Mathematics and Statistics, University of New Brunswick, 
Fredericton, NB, Canada E3B 5A3}

\begin{abstract}

Using a massless scalar field as a clock variable, the Legendre transform of a group field theory Lagrangian gives a relational Hamiltonian.  In the classical theory, it is natural to define `equal relational time' Poisson brackets, where `equal time' corresponds to equal values of the scalar field clock.  The quantum theory can then be defined by imposing `equal relational time' commutation relations for the fundamental operators of the theory, with the states being elements of a Fock space with their evolution determined by the relational Hamiltonian operator.  A particularly interesting type of states are condensates, as they may correspond to the cosmological sector of group field theory.  For the relational Hamiltonian considered in this paper, the coarse-grained dynamics of a simple family of condensate states agree exactly with the Friedmann equations in the classical limit, and also include quantum gravity corrections that ensure the big-bang singularity is replaced by a bounce.

\end{abstract}

\maketitle

\section{Introduction}
\label{s.intro}

In general relativity, the Hamiltonian is a constraint that generates gauge transformations corresponding to time-like diffeomorphisms, which include translations in time.  Therefore, in a quantum theory of gravity physical states are annihilated by the Hamiltonian constraint operator, and gauge-invariant observables (also known as Dirac observables) are necessarily invariant under time translations.  Because the observables in physical states are `frozen' in this sense, there is no immediate notion of dynamics, or of evolution in time.  This is known as the problem of time in quantum gravity.

One possible solution to the problem of time is to consider relational observables \cite{Rovelli:1990ph, Marolf:1994nz, Gambini:2000ht, Rovelli:2001bz, Dittrich:2005kc}.  These are observables where the first operator $\mO_1$ is evaluated at the `instant' that another observable $\mO_2$ playing the role of a relational clock has a specific value $\la_o$.  Then, by considering how the expectation value (or higher moments) of $\mO_1$ depends on the value of $\mO_2$, it is possible to speak of the `relational evolution' of $\mO_1$ with respect to $\mO_2$.

While the construction of relational observables is always possible for any operators $\mO_1$ and $\mO_2$, there is no guarantee that the resulting `relational evolution' is meaningful or can adequately address the problem of time.  Relational observables are typically most useful when $\mO_2$ corresponds to an observable that in the classical theory evolves monotonically in time, and so can be used as a good (monotonic) clock with respect to which the dynamics can be expressed.

Note that relational observables can be constructed in different ways \cite{Isham:1992ms}: (i) first isolating a clock variable $T$ before quantization and then using this clock variable to construct a quantum theory with a Schr\"odinger-like equation with time $T$ (this is the approach followed in this paper); (ii) first constructing a quantum theory following the Dirac programme and then finding a relational time observable in the resulting quantum theory with respect to which relational dynamics can be expressed; or (iii) if there is not a globally available internal time variable, after constructing a quantum theory following the Dirac programme it may nonetheless be possible to extract meaningful relational dynamics at least in some regimes.  While quantum theories of these different types will in general be inequivalent, it has recently been argued that, given a particular time variable, there exists a systematic quantum reduction map from a quantum theory of type (ii) (or type (iii) if one can find a good clock locally) to the type (i) form of the theory; in this sense, while quantum theories of type (ii) or (iii) are more general in that they are perspective-neutral, the (typically simpler) quantum theory of type (i) does in fact fully capture the physics relative to that particular reference clock \cite{Hoehn:2018aqt, Hoehn:2018whn}.

One simple example of a good relational clock is a massless scalar field: this has been used to define relational dynamics in, among others, canonical loop quantum gravity \cite{Domagala:2010bm} and loop quantum cosmology \cite{Ashtekar:2006wn}.  Other examples of matter fields used as relational clocks include dust fields \cite{Brown:1994py, Giesel:2007wn, Husain:2011tk} and Maxwell fields \cite{Pawlowski:2014fba}; clocks corresponding to geometric observables have been considered as well \cite{Marolf:1994wh, MartinBenito:2009qu}.

Relational observables have also been used in group field theory (GFT), a second-quantized reformulation of loop quantum gravity (LQG).  Specifically, relational dynamics were extracted from GFT condensate states.  As condensate states impose homogeneity in the sense that all quanta are in the same state, it has been conjectured that condensates could correspond to (at least part of) the cosmological sector of GFT \cite{Gielen:2013kla}.  For this reason, it is interesting to determine how the spatial volume of the space-time evolves with respect to the massless scalar field that is the matter content of the space-time.  These relational equations of motion can be translated into the Friedmann equations; they have the correct classical limit for a flat Friedmann-Lema\^itre-Robertson-Walker (FLRW) space-time minimally coupled to a massless scalar field, and they also include quantum gravity corrections that generically replace the big-bang singularity by a non-singular bounce \cite{Oriti:2016qtz}.

The successful use of relational observables to extract the dynamics of GFT condensate states suggests that it may be possible to use a relational framework more generally in group field theory.  In this paper, a relational Hamiltonian is derived for group field theory from the Legendre transformation of the theory's Lagrangian, using a matter field as the clock variable.

\section{Group Field Theory}
\label{s.gft}

One motivation to introduce group field theory comes from loop quantum gravity.  Spin foam models, a covariant approach to loop quantum gravity, gives a prescription to calculate transition amplitudes between an `initial' and a `final' quantum state using a sum-over-histories approach \cite{Perez:2012wv}.

A possible basis for the initial and final quantum states is given by spin-networks, i.e., graphs coloured by group elements.  It is possible to interpret a spin-network as a many-particle state, with each spin-network node and the `half-links' emanating from it corresponding to one (group-valued) `particle' or `quantum of geometry' \cite{Oriti:2013aqa}.  (These quanta can be joined into a spin-network by acting on the `node' particles with gluing operators in the form of projectors.)  Non-trivial dynamics occur when spin-network nodes meet and interact.  In this langauge, a spin foam model provides Feynman rules for the allowed interactions between spin-network nodes in the theory.

It is possible to construct a field theory with an action $S[\vp]$ such that the perturbative expansion of the partition function $Z = \int {\cal D}\vp \, e^{-S[\vp]}$ gives the Feynman rules for any spin foam model \cite{DePietri:1999bx, Reisenberger:2000zc}.  For this to work, the field $\vp$ must carry the same degrees of freedom as a spin-network node (and the `half-links' emanating from it), and therefore $\vp$ must be defined on a group manifold.  Such a theory is known as a group field theory (GFT).  Given the GFT action, it is then possible to study quantum gravity effects, not only in terms of the spin-foam amplitudes that arise perturbatively from the GFT partition function, but potentially also using other non-perturbative techniques.

Note that $S[\vp]$ appearing in the GFT partition function can be interpreted as an action, using a quantum field theory interpretation, or as an `energy' (although perhaps with a different physical interpretation) in the statistical weighting of different GFT excitations, using a statistical mechanical interpretation as proposed for example in \cite{Kotecha:2018gof}.  Here, following the quantum field theory interpretation, $S[\vp]$ is taken to be the GFT action.

The group field theory of interest here describes 4-dimensional Lorentzian gravity minimally coupled to a massless scalar field $\chi$ that will be used as a relational clock.  Specifically, this GFT is based on a real-valued field $\vp$ defined on $SU(2)^{\times 4} \times \mathbb{R}$, denoted by $\vp(g_1, g_2, g_3, g_4, \chi)$ with $g_i \in SU(2)$ and $\chi \in \mathbb{R}$; there exists an embedding of the $SU(2)$ group elements $g_i$ in $SL(2,\mathbb{C})$ that captures the Lorentzian geometric degrees of freedom \cite{Engle:2007wy}, while $\chi$ represents the matter content.  (Only four-valent spin-network nodes are allowed in this particular GFT.)  The field $\vp$ must satisfy the equation of motion
\be \label{gft-cl}
\f{\de S[\vp]}{\de \vp} = 0,
\ee
where $S[\vp]$ is the (classical) GFT action that generates the Feynman rules for the spin foam model (the form of the GFT action for this theory is given below), and $\vp$ must also satisfy the gauge-invariance condition of invariance under group multiplication from the right, $\vp(g_1, g_2, g_3, g_4) = \vp(g_1 h, g_2 h, g_3 h, g_4 h) ~~ \forall \,\, h \in SU(2)$.  (This is called the gauge-invariance condition since it encodes the gauge invariance satisfied by nodes in a spin-network.)  For more on GFT see, e.g., \cite{Oriti:2013aqa, Freidel:2005qe, Oriti:2006se}.

It is often convenient to use the Peter-Weyl theorem to express the GFT field as
\be \label{rep}
\vp(\vec g, \chi) = \!\! \sum_{j_i, m_i, n_i, \iota} \!\!\! \vp^{\vec \jmath, \iota}_{\vec m}(\chi) \,\,
\mI^{\vec \jmath, \iota}_{\vec n} \,\, \prod_{a=1}^4 \f{1}{d(j_a)} \, D^{j_a}_{m_a n_a}\!(g_a),
\ee
where $d(j_a) = 2j_a + 1$, the $D^j_{mn}(g)$ are the Wigner matrices, and the intertwiners $\mI$ arise due to the invariance under group multiplication from the right in the GFT field.  Note that since the Wigner matrices are complex-valued, the $\vp^{\vec \jmath, \iota}_{\vec m}(\chi)$ are also complex.  Here $\vp(\vec g, \chi) \equiv \vp(g_1, g_2, g_3, g_4, \chi)$ and it is understood that there are 4 $j$ and 4 $m$ labels in each of $\vp^{\vec \jmath, \iota}_{\vec m}$ and $\mI^{\vec \jmath, \iota}_{\vec n}$, for example
\be
\vp^{\vec \jmath, \iota}_{\vec m}(\chi) \equiv \vp^{j_1, j_2, j_3, j_4, \iota}_{m_1, m_2, m_3, m_4}(\chi),
\ee
and $\iota$ labels the possible intertwiners for each set $(j_1, j_2, j_3, j_4)$.  Each pair of $j_i, m_i$ lives on one of the four half-links leaving the node, while the intertwiner $\iota$ lives at the node.  It is helpful to choose a basis of intertwiners that diagonalizes the action of the loop quantum gravity volume operator on a spin-network node; then in the quantum theory, quanta corresponding to $\vp^{\vec \jmath, \iota}_{\vec m}$ will carry a definite volume $V_{\vec \jmath, \iota}$ given by the corresponding eigenvalue of the LQG volume operator.

The GFT action is chosen with an eye on the resulting quantum theory.  Specifically, the kinetic and potential terms in the action are chosen so that the Feynman rules for the GFT give the edge and vertex amplitudes, respectively, in spin foam models.  There are two key inputs that will be used here: (i) the edge amplitude in the spin foam models of interest here carries a Kronecker delta in the $j_i, m_i$ and $\iota$ labels \cite{Engle:2007wy}, and (ii) for spin foam models with a scalar field as matter content, the vertex amplitude is expected to provide a product of Dirac delta functions between the values of $\chi$ in the GFT fields $\vp$ that meet at the vertex, implying that the interaction is local in $\chi$ \cite{Oriti:2016qtz, Li:2017uao}.  These properties can be used to restrict the form of the GFT action.  In addition, in simplicial gravity the interaction term is fifth-order in the GFT field, $\sim \vp^5$.  For concreteness, this is the form assumed here for the GFT action, but this can easily be generalized to allow other types of interactions.  While these general arguments provide some restrictions on the possible terms that could appear in the GFT action if one wishes to recover general relativity in a classical limit, it is not yet known exactly which terms are required.  Because of this uncertainty, in the following I will consider only what are expected to be the leading order terms in the GFT action, and I will leave the form of the couplings to be as general as possible.

Given the above, the GFT action is
\be
S[\vp] = K[\vp] - V[\vp],
\ee
where the kinetic term, whose inverse is the spin foam propagator, has the form
\be
K[\vp] = \f{1}{2} \int \! \dd \chi \dd \t\chi \! \sum_{j, m, \iota} \! \vp^{\vec \jmath, \iota}_{\vec m}(\chi) \,
\mK^{\vec \jmath, \iota}_{\vec m}(\chi, \t\chi) \, \vp^{\vec \jmath, \iota}_{\vec m}(\t\chi),
\ee
where the sum is over four $j$ labels, four $m$ labels, and one intertwiner label, while the interaction term is
\be
V[\vp] = \! \f{1}{5} \int \! \dd\chi \sum_{j,m,\iota} \! \mV^{j_i, m_i, \iota_i}(\chi) \! \prod_{a=1}^5 \vp^{\vec \jmath_a, \iota_a}_{\vec m_a}(\chi),
\ee
the sum is over 20 $j$ labels, 20 $m$ labels and 5 intertwiners $\iota$, each $\vp$ contributing 4 $j$, 4 $m$ and one $\iota$.  Here $\mK^{\vec \jmath, \iota}_{\vec m}(\chi, \t\chi)$ and $\mV^{j_i, m_i, \iota_i}(\chi)$ are as yet unspecified functions of the $j, m, \iota$ and $\chi$ labels; determining which choices give a good quantum gravity theory remains an important open problem.  For the action to be real, the kinetic term must satisfy $\bar\mK^{\vec \jmath, \iota}_{\vec m}(\chi, \t\chi) = \mK^{\vec \jmath, \iota}_{-\vec m}(\chi, \t\chi)$, and a similar condition (whose exact form depends on the combinatorial structure of the potential) is required for $\mV^{j_i, m_i, \iota_i}(\chi)$.  To work in a framework for GFT that is as general as possible, other than imposing some simple symmetries (described next) $\mK^{\vec \jmath, \iota}_{\vec m}(\chi, \t\chi)$ and $\mV^{j_i, m_i, \iota_i}(\chi)$ will be left free.

The specific form of the GFT action will depend on the matter content.  For the case of $\chi$ being a massless scalar field, it is reasonable to assume that the kernel in the interaction term, $\mV^{j_i,m_i,\iota_i}(\chi)$, is independent of $\chi$,
\be
V[\vp] = \! \f{1}{5} \int \! \dd\chi \sum_{j,m,\iota} \! \mV^{j_i, m_i, \iota_i} \! \prod_{a=1}^5 \vp^{\vec \jmath_a, \iota_a}_{\vec m_a}(\chi).
\ee
Results obtained for vacuum Lorentzian gravity suggest the kernel $\mV$ should involve a $15j$ symbol \cite{Engle:2007wy}.

Finally, the symmetries of the action of a massless scalar field minimally coupled to gravity (invariance under $\chi \to -\chi$ and $\chi \to \chi + \chi_o$) suggest that $\mK$ should have the form \cite{Oriti:2016qtz}
\be
\mK^{\vec \jmath, \iota}_{\vec m}(\chi, \t\chi) = \mK^{\vec \jmath, \iota}_{\vec m} \big((\chi - \t\chi)^2 \big).
\ee

Since the kinetic term in the action is non-local in $\chi$ (in that it consists of two integrals over $\chi$ and $\t\chi$ rather than one integral over $\chi$), it is convenient to rewrite it in terms of a derivative expansion in $\chi$.  Defining $\t\chi = \chi + \de\chi$, and performing a Taylor expansion in $\vp^{\vec \jmath, \iota}_{\vec m}(\chi+\de\chi)$ around $\chi$ gives
\be \label{exp-k}
K = \f{1}{2} \sum_{n=0}^\infty \int \dd\chi \sum_{j, m, \iota} \vp^{\vec \jmath, \iota}_{\vec m}(\chi)
\, \mK^{(2n)}_{\vec \jmath, \vec m, \iota} ~ \partial_\chi^{2n} \vp^{\vec \jmath, \iota}_{\vec m}(\chi),
\ee
with
\be
\mK^{(2n)}_{\vec \jmath, \vec m, \iota} = \int \dd(\de\chi) \f{(\de\chi)^{2n}}{(2n)!} \mK^{\vec \jmath, \iota}_{\vec m}(\de\chi^2).
\ee

Note that since the scalar field $\chi$ has dimensions of mass and assuming that $\vp$ is dimensionless, $\mK^{(2n)}_{\vec \jmath, \vec m, \iota}$ must scale as $M_{Pl}^{2n-1}$ as the Planck mass is the only dimensionful constant in the theory.  Since higher order derivatives are suppressed by higher powers of $\hbar$, the leading order terms in $K$ are $n=0, 1$.

Therefore, the leading order contributions to the GFT action for gravity minimally coupled to a massless scalar field $\chi$ has the form
\be \label{gft-s}
S[\vp] = \int \dd\chi \, \mL[\vp],
\ee
with
\begin{align} \label{gft-l}
\mL[\vp] = & \, - \f{1}{2} \sum_{j,m,\iota} \Big( \partial_\chi \vp^{\vec \jmath, \iota}_{\vec m}(\chi) \Big) \, \mK^{(2)}_{\vec \jmath, \vec m, \iota} ~ \Big( \partial_\chi \vp^{\vec \jmath, \iota}_{\vec m}(\chi) \Big) \nn \\ & ~
+ \f{1}{2} \sum_{j,m,\iota} \vp^{\vec \jmath, \iota}_{\vec m}(\chi) \mK^{(0)}_{\vec \jmath, \vec m, \iota} ~ \vp^{\vec \jmath, \iota}_{\vec m}(\chi)
- U[\vp],
\end{align}
where the first term in \eqref{gft-l} was integrated by parts and
\be \label{gft-pot}
U[\vp] = \f{1}{5} \sum_{j,m,\iota} \! \mV^{j_i, m_i, \iota_i} \! \prod_{a=1}^5 \vp^{\vec \jmath_a, \iota_a}_{\vec m_a}(\chi).
\ee
As explained above, the sum in the kinetic terms is over 4 $j_i$, 4 $m_i$ and one $\iota$ label, while the sum in the potential $U[\vp]$ is over 20 $j_i$, 20 $m_i$ and 5 $\iota$ labels.  Note that while the GFT action is non-local in the group elements, to leading order in the derivative expansion with respect to $\chi$, the GFT action is local in $\chi$.

This GFT action has the same form as the action for a classical field theory (although the base manifold is $SU(2)^{\times 4} \times \mathbb{R}$, not the $\mathbb{R}^4$ of space-time), where $\chi$ now plays the role of time.  The equation of motion for the GFT field $\vp$ can be derived directly from \eqref{gft-cl}, or equivalently as the Euler-Lagrange equation for the GFT action $S[\vp]$ using $\chi$ as a time variable, with the result
\be \label{gft-el}
\mK^{(2)}_{\vec \jmath, \vec m, \iota} ~ \partial_\chi^2 \vp^{\vec \jmath, \iota}_{\vec m}(\chi)
+ \mK^{(0)}_{\vec \jmath, \vec m, \iota} ~ \vp^{\vec \jmath, \iota}_{\vec m}(\chi) - \f{\de U[\vp]}{\de \vp^{\vec \jmath,\iota}_{\vec m}} = 0,
\ee
which holds for each $\vec \jmath, \vec m, \iota$.  The explicit expression for $\de U / \de \vp$ can be calculated from \eqref{gft-pot}.

While the GFT Lagrangian \eqref{gft-l} was constructed so that the perturbative expansion of the partition function reproduces the Feynman rules of a spin foam model, it is now possible to use the (classical) GFT action to develop other complementary formulations of the quantum theory.

\section{Legendre Transform}
\label{s.tr}

Given the GFT action \eqref{gft-s}, it is straightforward to calculate its Legendre transform with respect to $\chi$.

First, the momentum $\pi$ conjugate to the GFT field $\vp$ is
\be \label{momentum}
\pi^{\vec \jmath,\iota}_{\vec m}(\chi) = \f{\delta \mL}{\delta \Big(\partial_\chi \vp^{\vec \jmath,\iota}_{\vec m} \Big)}
= - \mK^{(2)}_{\vec \jmath,\vec m,\iota} \partial_\chi \vp^{\vec \jmath,\iota}_{\vec m}(\chi),
\ee
since the potential is independent of $\partial_\chi \vp$.  Then, the Legendre transform of the Lagrangian $\mL$ with respect to $\chi$ gives the `relational Hamiltonian'
\begin{align} \label{gft-h}
\!\! \mH[\vp] & = \Big( \sum_{j,m,\iota} \pi^{\vec \jmath,\iota}_{\vec m}(\chi) \,\, \partial_\chi \vp^{\vec \jmath,\iota}_{\vec m}(\chi) \Big) - \mL \nn \\ &
= - \sum_{j,m,\iota} \left[ \f{\pi^{\vec \jmath, \iota}_{\vec m}(\chi)^2} {2 \, \mK^{(2)}_{\vec \jmath, \vec m, \iota}}
+ \mK^{(0)}_{\vec \jmath, \vec m, \iota} \, \f{\vp^{\vec \jmath, \iota}_{\vec m}(\chi)^2}{2} \right] \!
+ U[\vp].
\end{align}
Defining the `equal relational time' Poisson brackets
\be \label{pb}
\{ \vp^{\vec \jmath_1, \iota_1}_{\vec m_1}(\chi), \pi^{\vec \jmath_2,\iota_2}_{\vec m_2}(\chi) \} = \de^{\vec \jmath_1,\vec \jmath_2} \de_{\vec m_1, \vec m_2} \de^{\iota_1, \iota_2},
\ee
the equations of motion for any ($\chi$-independent) observable $\mO$ in the GFT can be derived from
\be
\f{d \mO}{d\chi} = \{\mO, \mH\}.
\ee
In particular, the equation of motion for $\partial_\chi \vp$ gives \eqref{momentum}, and this combined with the equation of motion for $\partial_\chi \pi$ gives precisely the Euler-Lagrange equation \eqref{gft-el} derived from the GFT Lagrangian $\mL[\vp]$.

Note that the `equal relational time' Poisson brackets defined in \eqref{pb} are analogous (though not identical) to the Poisson brackets posited in \cite{Adjei:2017bfm} (see also Sec.~V in \cite{Kotecha:2018gof}).  On the other hand, they are quite different to the timeless Poisson brackets typically defined in GFT \cite{Oriti:2013aqa}.

\section{Quantum Theory}
\label{s.qt}

An advantage of having a (relational) Hamiltonian for GFT is that it is possible to perform a canonical quantization of the theory, treating $\chi$ as a classical time variable.

The first step is to replace the equal time Poisson bracket for the basic GFT field $\vp$ and its momentum $\pi$ by commutation relations for the corresponding operators,
\be \label{comm}
[ \, \h\vp^{\vec \jmath_1, \iota_1}_{\vec m_1}, \, \h \pi^{\vec \jmath_2,\iota_2}_{\vec m_2} \, ] = i \hbar \, \de^{\vec \jmath_1,\vec \jmath_2} \de_{\vec m_1, \vec m_2} \de^{\iota_1, \iota_2}.
\ee
The above expression is given in the Schr\"odinger picture, in the Heisenberg picture \eqref{comm} would be given in terms of `equal relational time' commutation relations.  As usual, these operators can formally be represented by
\begin{gather}
\h\vp^{\vec \jmath, \iota}_{\vec m} \Psi[\vp] = \vp^{\vec \jmath, \iota}_{\vec m} \, \Psi[\vp], \\
\h\pi^{\vec \jmath, \iota}_{\vec m} \Psi[\vp] = -i\hbar \, \f{\de \, \Psi[\vp]}{\de \vp^{\vec \jmath, \iota}_{\vec m}}.
\end{gather}

The relational Schr\"odinger equation for GFT wave functionals is
\be \label{schr}
\h \mH \, \Psi = i \hbar \, \f{d \Psi}{d\chi},
\ee
with the Hamiltonian operator
\be \label{qm-ham}
\h \mH = - \sum_{j,m,\iota} \left[ \f{\h \pi^{\vec \jmath, \iota}_{\vec m}(\chi)^2} {2 \, \mK^{(2)}_{\vec \jmath, \vec m, \iota}}
+ \mK^{(0)}_{\vec \jmath, \vec m, \iota} \, \f{\h \vp^{\vec \jmath, \iota}_{\vec m}(\chi)^2}{2} \right] \!
+ U[\h \vp].
\ee

Given that the Hamiltonian contains terms that are quadratic in the field and its momentum, it is convenient to introduce creation and annihilation operators corresponding respectively to
\begin{gather}
\h a_{\vec \jmath, \vec m, \iota}^\dag = \f{1}{\sqrt{2 \hbar \, \omega^{\vec \jmath, \iota}_{\vec m} \, }}
\left( \omega^{\vec \jmath, \iota}_{\vec m} \, \h \vp^{\vec \jmath, \iota}_{\vec m} - i \, \pi^{\vec \jmath, \iota}_{\vec m} \right), \\
\h a_{\vec \jmath, \vec m, \iota} = \f{1}{\sqrt{2 \hbar \, \omega^{\vec \jmath, \iota}_{\vec m} \, }}
\left( \omega^{\vec \jmath, \iota}_{\vec m} \, \h \vp^{\vec \jmath, \iota}_{\vec m} + i \, \pi^{\vec \jmath, \iota}_{\vec m} \right),
\end{gather}
with
\be
\omega^{\vec \jmath, \iota}_{\vec m} = \sqrt{|\mK^{(0)}_{\vec \jmath, \vec m, \iota} \, \mK^{(2)}_{\vec \jmath, \vec m, \iota}|}.
\ee
(Recall that $\mK^{(0)}_{\vec \jmath, \vec m, \iota}$ and $\mK^{(2)}_{\vec \jmath, \vec m, \iota}$ may be positive or negative, depending on the specific form of the GFT action.)  As usual, the creation and annihilation operators satisfy the commutation relations
\be \label{comm-a}
[ \, \h a_{\vec \jmath_1, \vec m_1, \iota_1}, \, \h a_{\vec \jmath_2, \vec m_2, \iota_2}^\dag \, ] = \de_{\vec \jmath_1,\vec \jmath_2} \de_{\vec m_1, \vec m_2} \de_{\iota_1, \iota_2}.
\ee

Since the ladder operators satisfy the commutation relations \eqref{comm-a}, GFT states live in the Fock space
\be
{\cal F} = \bigoplus_{n=0}^\infty S H^{\otimes n},
\ee
where $H$ is the `single-particle' Hilbert space for $SU(2)^{\otimes4}/SU(2)$ and $S$ denotes the symmetrization of $H^{\otimes n}$ required by the bosonic statistics of the theory.  As usual, the GFT Fock `vacuum' state $|0\ket$ is defined as the state annihilated by all $\h a_{\vec \jmath, \vec m, \iota}$ operators.  Note that $|0\ket$ is not necessarily an eigenstate of the Hamiltonian operator (even in the absence of a potential $U$) depending on the signs of $\mK^{(0)}_{\vec \jmath, \vec m, \iota}$ and $\mK^{(2)}_{\vec \jmath, \vec m, \iota}$.

The creation and annihilation operators make it possible to speak of quanta of geometry: the $\h a_{\vec \jmath, \vec m, \iota}^\dag$ create quanta of geometry, i.e., spin-network nodes labeled with $\vec\jmath, \vec m, \iota$.  So a spin-network with $N$ nodes can be constructed by acting on the GFT Fock vacuum with $N$ creation operators; connectivity between neighbouring nodes is imposed by projection operators acting on the two half-links that are to be connected, just as in the non-deparametrized theory \cite{Oriti:2013aqa}.  In this reformulation of LQG (for gravity minimally coupled to a massless scalar field), the physical Hilbert space can be understood as the space of states that live in the GFT Fock space and satisfy the relational Schr\"odinger equation, which determines the relational quantum gravity dynamics.  (It may also be possible to use other, non-Fock, representations of states \cite{Kegeles:2017ems}.)

The relational Hamiltonian can be expressed in terms of the creation and annihilation operators.  If $\mK^{(0)}_{\vec \jmath, \vec m, \iota}$ and $\mK^{(2)}_{\vec \jmath, \vec m, \iota}$ have the same sign, then
\be \label{ham-1}
\h \mH = \hbar \sum_{j, m, \iota} \!\! M_{\vec \jmath, \vec m, \iota} \left( \h a_{\vec \jmath, \vec m, \iota}^\dag \h a_{\vec \jmath, \vec m, \iota} + \f{1}{2} \right) + U[\h\vp],
\ee
while if they have opposite signs
\be \label{ham-2}
\h \mH = \f{\hbar}{2} \sum_{j, m, \iota} \! M_{\vec \jmath, \vec m, \iota} \Big( (\h a_{\vec \jmath, \vec m, \iota}^\dag)^2 + \h a_{\vec \jmath, \vec m, \iota}^2 \Big) + U[\h\vp].
\ee
In both cases
\be
M_{\vec \jmath, \vec m, \iota} = \pm \sqrt{\left|\f{\mK^{(0)}_{\vec \jmath, \vec m, \iota}}{\mK^{(2)}_{\vec \jmath, \vec m, \iota}}\right|},
\ee
where the signs of $\mK^{(0)}_{\vec \jmath, \vec m, \iota}$ and $\mK^{(2)}_{\vec \jmath, \vec m, \iota}$ determine the overall sign of $M_{\vec \jmath, \vec m, \iota}$ by ensuring that the overall signs agree between \eqref{qm-ham} and either \eqref{ham-1} or \eqref{ham-2}.

The exact form of the $M_{\vec \jmath, \vec m, \iota}$ is to be determined by the GFT action for quantum gravity (whose form, as discussed in Sec.~\ref{s.gft}, has been partially constrained by some symmetries but is not yet exactly known).  Perhaps the simplest possibility would be $M_{\vec \jmath, \vec m, \iota} = M$, but in models of this type quantum fluctuations typically generate a Laplace-Beltrami operator acting on each $SU(2)$ element term in $\mK^{(0)}$ \cite{Geloun:2011cy}; this suggests the form $M_{\vec \jmath, \vec m, \iota} = \pm [a + b \sum_{i=1}^4 j_i(j_i+1)]^{1/2}$, with $a$ and $b$ the free parameters of the theory.  However, in the absence of a clear derivation of the correct form for $M_{\vec \jmath, \vec m, \iota}$, it appears safer to avoid making any specific choices at this point and to leave $M_{\vec \jmath, \vec m, \iota}$ in its most general form possible.

Note that it is only for the Hamiltonian \eqref{ham-1} that the state $|0\ket$ is an eigenstate of the Hamiltonian.  Of course, it is also possible that the two $\mK$ have the same sign for some quantum labels $\vec \jmath, \vec m, \iota$, and a different sign for others.  As soon as one pair of $\mK^{(0)}$ and $\mK^{(2)}$ have opposite signs, $|0\ket$ is not an eigenstate of the Hamiltonian.

If the Hamiltonian has at least one term of the form \eqref{ham-2}, then since the Fock vacuum state $|0\ket$ is not an eigenstate of $\h \mH$, $|0\ket$ cannot be the vacuum state of the relational Hamiltonian, even approximately (and this is the case even if the interaction term vanishes).  In this case, the `no-space' state $|0\ket$ is unstable and if initial conditions are chosen such that at some initial time $\chi_o$ the state is $|\Psi(\chi_o)\ket = |0\ket$, the state will evolve away from $|0\ket$ following
\be
|\Psi(\chi)\ket = e^{-i \h \mH (\chi - \chi_o) / \hbar} |\Psi(\chi_o)\ket.
\ee
In this case, for the $\vec\jmath, \vec m, \iota$ for which the relational Hamiltonian has the form \eqref{ham-2}, the potential (neglecting $U$) is an inverted harmonic oscillator with an unstable fixed point that quantum fluctuations will push the system away from, at which point the system will `roll down' the potential.  As a result, an initial state $|0\ket$ will rapidly become excited, and since each quantum of geometry contributes some spatial volume, this state can be understood as a sort of expanding universe.  This provides a realization of geometrogenesis: the instability of the no-space state ensures that geometric excitations arise.  But more precisely, this is a type of bounce that goes through the zero-volume state since the evolution in the other direction of (relational) time is identical, due to the discrete symmetry $\chi \to -\chi$ in the GFT action and the invariance of $|0\ket$ under relational time reversal.  Quantum dynamics do not break down at the zero-volume state but simply evolve past it; these dynamics are studied in more detail towards the end of Sec.~\ref{s.cond}.

The instability in the Hamiltonian is important for cosmology: the universe can expand indefinitely.  Quanta contribute volume to the space-time and having a Hamiltonian where the GFT field can escape to infinity is one way to allow for an unending expansion of the universe.  Based on this argument, from a cosmological perspective the relational Hamiltonian of the form \eqref{ham-2} appears to be the most interesting one.  (Of course, the same Hamiltonian should be used for all states in GFT, no matter the space-time they may correspond to, but the cosmological sector suggests which choice for $\h \mH$ may be the correct one.)  It is interesting that an essentially identical Hamiltonian was proposed for the toy model considered in \cite{Adjei:2017bfm}, based only on cosmological considerations; here it arises from the relational quantization of the GFT action \eqref{gft-s}.

Finally, the quantum theory has been defined here in the Schr\"odinger picture, with the quantum state evolving in relational time with respect to \eqref{schr} while the basic operators, like the creation and annihilation operators, are independent of time.  It is also possible to work in the Heisenberg picture, where states are independent of relational time while operators evolve as
\be
\h \mO(\chi) = e^{i \h\mH (\chi-\chi_o)/\hbar} \, \h\mO(\chi_o) \, e^{-i \h\mH (\chi-\chi_o)/\hbar}.
\ee
The existence of the unitary relational time evolution operator $U(\chi,\chi_o) = \exp[-i \h \mH (\chi - \chi_o) / \hbar]$ ensures that the Heisenberg and Schr\"odinger pictures are equivalent; in this way it is similar to some other (simpler) constrained systems where the relational Heisenberg and Schr\"odinger pictures have also been shown to be equivalent \cite{Olmedo:2016zlp}.

\section{Cosmology}
\label{s.cond}

In the context of GFT, it has been argued that the simplest condensate states may correspond to (at least a portion of) the cosmological sector of the theory, since condensate states impose homogeneity at a microscopic and macroscopic level \cite{Gielen:2013kla}.

At this time, the potential correspondence between GFT condensate states and cosmology remains a conjecture, but previous studies based on this idea for a non-deparametrized GFT found that the emergent coarse-grained dynamics of simple condensate states correspond to the Friedmann equations of the flat FLRW space-time with the correct classical limit, and also include quantum gravity corrections that become important only in the Planck regime \cite{Oriti:2016qtz, Oriti:2016ueo}.  Based on these results, it appears promising to further explore this potential correspondence in condensate states of a GFT with a relational Hamiltonian.

More specifically, the conjectured correspondence between GFT condensates and cosmology suggests that in the limit that connectivity between spin-network nodes is ignored and that the interaction term $U$ is negligible in the Hamiltonian, the flat FLRW space-time is expected to emerge from a suitable coarse-graining of a coherent state of the GFT creation operator.  (These two assumptions are related as the interaction term creates correlations between quanta of geometry, and so generates connectivity.)

The assumption that connectivity can be ignored for the flat FLRW space-time is based on two arguments: (i) there is no need to encode spatial curvature in the connectivity, and (ii) the main observable of interest---namely the spatial volume---is independent of the connectivity structure between the quanta of geometry.  In addition, the interaction term in the Hamiltonian is expected to be negligible when there are sufficiently few quanta so that (the expectation value of) the interaction term is negligible compared to that of the kinetic term in the Hamiltonian.  For more on these approximations, see \cite{Oriti:2016qtz, Oriti:2016ueo}.  If the interaction term is not negligible, connectivity may become important in which case a simple coherent state may not be sufficient to capture the physics of interest.  The results derived in this section only hold in the regime where connectivity and the interaction term can safely be neglected.

Coherent states, in the Heisenberg picture, have the form
\be \label{def-sigma}
|\sigma\ket = e^{-\| \sigma \|^2/2} \, e^{\sum_{j,m,\iota} \sigma^{\vec \jmath,\iota}_{\vec m} \, a_{\vec \jmath, \vec m, \iota}(\chi_o)^\dag} |0\ket,
\ee
where $\| \sigma \|^2 = \sum_{j,m,\iota} |\sigma^{\vec \jmath,\iota}_{\vec m}|^2$ and $\chi_o$ is an arbitrary initial relational time.  Here the (complex-valued) condensate wave function $\sigma^{\vec \jmath,\iota}_{\vec m}$ determines the weighting of the creation operators for different excitations in the exponential.  An important property of the states \eqref{def-sigma} is
\be \label{exp-a}
\bra \sigma| a_{\vec \jmath, \vec m, \iota}(\chi_o) |\sigma \ket = \sigma^{\vec \jmath,\iota}_{\vec m}.
\ee

In cosmology, one of the essential geometric quantities is the spatial volume.  The spatial volume operator in GFT is
\be
\h V = \sum_{j,m,\iota} V_{\vec \jmath, \iota} a_{\vec \jmath, \vec m, \iota}^\dag a_{\vec \jmath, \vec m, \iota},
\ee
where $V_{\vec \jmath, \iota}$ denotes the eigenvalue of the LQG volume operator acting on a spin-network node with the quantum numbers $\vec \jmath, \iota$ (recall the intertwiners have been chosen to diagonalize the LQG volume operator).  From \eqref{exp-a}, the expectation value of the spatial volume at the initial relational time $\chi_o$ for the state $|\sigma\ket$ is
\be \label{exp-v}
\bra \h V(\chi_o) \ket_\sigma = \sum_{j,m,\iota} V_{\vec \jmath, \iota} |\sigma^{\vec \jmath,\iota}_{\vec m}|^2.
\ee

One of the key questions is to understand how $\bra V \ket_\sigma$ evolves.  At the initial relational time $\chi_o$,
\begin{align} \label{da}
\bra \f{\dd \h a_{\vec \jmath, \vec m, \iota}}{\dd\chi} \ket_\sigma &= \f{1}{i\hbar} \bra [a_{\vec \jmath, \vec m, \iota}, \h\mH] \ket_\sigma \nn \\ &
= - i M_{\vec \jmath, \vec m, \iota} \bra a_{\vec \jmath, \vec m, \iota}^\dag \ket_\sigma = - i M_{\vec \jmath, \vec m, \iota} \, \b\sigma^{\vec \jmath,\iota}_{\vec m},
\end{align}
assuming that the Hamiltonian has the form \eqref{ham-2} (following the arguments towards the end of Sec.~\ref{s.qt}) and dropping the potential $U[\h \vp]$ which has been assumed to be negligible.

For systems other than a simple harmonic oscillator, the coherent states \eqref{def-sigma} cannot be expected to be exact solutions of the quantum dynamics, but in some situations, especially when the interaction term is negligible as has been assumed here, they can provide a good approximate solution.  This suggests that it is a reasonable approximation to assume that the relations \eqref{exp-v} and \eqref{da} hold for some interval of (relational) time, not just at $\chi=\chi_o$.  In this approximation, the expectation value for the spatial volume is given by
\be
\bra \h V(\chi) \ket_\sigma = \sum_{j,m,\iota} V_{\vec \jmath, \iota} |\bra a_{\vec \jmath, \vec m, \iota}(\chi)\ket_\sigma|^2,
\ee
while the dynamics follow from \eqref{da},
\be \label{eom-a}
\f{\dd \bra a_{\vec \jmath, \vec m, \iota}(\chi)\ket_\sigma}{\dd\chi}
= - i M_{\vec \jmath, \vec m, \iota} \, \bra a_{\vec \jmath, \vec m, \iota}(\chi) \ket_\sigma^*,
\ee
with ${}^*$ indicating complex conjugation.

Denoting the complex expectation value
\be
\bra \h a_{\vec \jmath, \vec m, \iota}(\chi) \ket_\sigma = \rho_{\vec \jmath, \vec m, \iota}(\chi) e^{i \theta_{\vec \jmath, \vec m, \iota}(\chi)},
\ee
the initial conditions are given by
\be
\rho_{\vec \jmath, \vec m, \iota}(\chi_o) e^{i \theta_{\vec \jmath, \vec m, \iota}(\chi_o)} = \sigma^{\vec \jmath,\iota}_{\vec m},
\ee
and the dynamics, given the assumptions described above, are
\begin{gather}
\f{\dd\rho_{\vec \jmath, \vec m, \iota}}{\dd\chi} = - \, M_{\vec \jmath, \vec m, \iota} \, \rho_{\vec \jmath, \vec m, \iota} \sin 2\theta_{\vec \jmath, \vec m, \iota}, \\
\f{\dd\theta_{\vec \jmath, \vec m, \iota}}{\dd\chi} = - \, M_{\vec \jmath, \vec m, \iota} \cos 2\theta_{\vec \jmath, \vec m, \iota}.
\end{gather}
The second differential equation can be solved by separation of variables, with the result
\be \label{theta-sol}
\sin2\theta_{\vec \jmath, \vec m, \iota} = - \tanh \left( 2M_{\vec \jmath, \vec m, \iota} (\chi - \t\chi_{\vec \jmath, \vec m, \iota}^o) \right),
\ee
where $\t\chi_{\vec \jmath, \vec m, \iota}^o$ is a constant of integration determined by $\theta_{\vec \jmath, \vec m, \iota}(\chi_o)$.  From this, it is straightforward to integrate the equation of motion for $\rho_{\vec \jmath, \vec m, \iota}$, giving
\be \label{rho-sol}
\rho_{\vec \jmath, \vec m, \iota}^2 = A_{\vec \jmath, \vec m, \iota} \cosh \left( 2M_{\vec \jmath, \vec m, \iota} (\chi - \t\chi_{\vec \jmath, \vec m, \iota}^o) \right),
\ee
with $A_{\vec \jmath, \vec m, \iota} \ge 0$ another constant of integration fixed by the initial condition $\rho_{\vec \jmath, \vec m, \iota}(\chi_o)$.

As a result, the expectation value for the spatial volume as a function of relational time is
\be \label{sol-v}
\! \bra \h V \ket_\sigma = \!\! \sum_{j,m,\iota} \! V_{\vec \jmath, \iota} A_{\vec \jmath, \vec m, \iota} \cosh \! \Big( 2M_{\vec \jmath, \vec m, \iota} (\chi - \t\chi_{\vec \jmath, \vec m, \iota}^o) \! \Big).
\ee
This clearly shows that, so long as one $A_{\vec \jmath, \vec m, \iota}$ is non-zero (which is true so long as at least one $\sigma^{\vec \jmath,\iota}_{\vec m} \neq 0$), the volume never vanishes.  Further, at very early and very late relational times, the volume becomes arbitrarily large.  As a result, the cosmology that emerges from coherent states in this relational GFT theory has a non-singular bounce with a large space-time on either side.

It is also possible to extract information about the matter sector: in the relational framework developed here, \eqref{schr} shows that the momentum of the massless scalar field (not to be confused with the momentum of the GFT field) is given by $\h \pi_\chi = i\hbar \partial_\chi = \h\mH$, so neglecting the potential term $U$ in the relational Hamiltonian,
\be
\h\pi_\chi = \frac{\hbar}{2} \sum_{j, m, \iota} M_{\vec \jmath, \vec m, \iota} \Big( (\h a_{\vec \jmath, \vec m, \iota}^\dag)^2 + \h a_{\vec \jmath, \vec m, \iota}^2 \Big).
\ee
Evaluated on the condensate state, this gives
\begin{align} \label{exp-pi}
\bra \pi_\chi \ket_\sigma &= \hbar \sum_{j, m, \iota} M_{\vec \jmath, \vec m, \iota} \rho_{\vec \jmath, \vec m, \iota} \cos 2\theta_{\vec \jmath, \vec m, \iota} \nn \\ &
= \hbar \sum_{j, m, \iota} M_{\vec \jmath, \vec m, \iota} A_{\vec \jmath, \vec m, \iota},
\end{align}
where the second equality is obtained by rewriting $\cos 2\theta = \sqrt{1 - \sin^2 2\theta}$ and using \eqref{theta-sol} and \eqref{rho-sol}.  (For a massless scalar field, the overall sign of $\pi_\chi$ is unimportant, so it is enough to take the positive root when solving for $\cos 2\theta$.)  Note that $\bra \pi_\chi \ket_\sigma$ is a constant of the motion, as should be expected: $\pi_\chi$ is a constant of motion for a free massless scalar field in an FLRW space-time.

Finally, consider the case where $A_{\vec \jmath, \vec m, \iota}$ does vanish for some $\vec \jmath, \vec m, \iota$.  In this case, the state is in the `particle vacuum' for that particular mode, which is unstable if the contribution to the Hamiltonian for these values of $\vec \jmath, \vec m, \iota$ is of the form \eqref{ham-2}.  Since the dynamics for each set of $\vec\jmath, \vec m, \iota$ decouple in the limit that $U$ is negligible, it is possible to determine the contribution for each $\vec\jmath, \vec m, \iota$ to the total volume separately, and sum over all $\vec\jmath, \vec m, \iota$ at the end.  For each $\vec \jmath, \vec m, \iota$ that has a vanishing $A_{\vec \jmath, \vec m, \iota}$, their contribution to the volume (denoted by $\tilde V_{\vec\jmath, \vec m, \iota}$) will evolve as
\be \label{vac-vol}
\bra \tilde V_{\vec\jmath, \vec m, \iota}(\chi) \ket = \bra 0 | \h S^\dag V_{\vec\jmath, \iota} \h a_{\vec\jmath, \vec m, \iota}(\chi_o)^\dag \h a_{\vec\jmath, \vec m, \iota}(\chi_o) \h S | 0 \ket,
\ee
where here in an abuse of notation $|0\ket$ is the zero-particle state for the mode labeled by $\vec\jmath, \vec m, \iota$ that is annihilated by $a_{\vec\jmath, \vec m, \iota}(\chi_o)$, and since the Hamiltonian for these values of $\vec \jmath, \vec m, \iota$ has the form \eqref{ham-2}, the unitary time evolution of the system is a squeezing operator:
\be
\! \h S = \exp \left( -\frac{i M_{\vec \jmath, \vec m, \iota}}{2} \Big( (\h a_{\vec \jmath, \vec m, \iota}^\dag)^2 + \h a_{\vec \jmath, \vec m, \iota}^2 \Big) (\chi - \chi_o) \right).
\ee

Using the Baker-Campbell-Hausdorff equation (see, e.g., \cite{Adjei:2017bfm} for intermediate steps), \eqref{vac-vol} gives
\be \label{vj0}
\bra \tilde V_{\vec\jmath, \vec m, \iota}(\chi) \ket = V_{\vec\jmath, \iota} \sinh^2 \Big(M_{\vec \jmath, \vec m, \iota}(\chi - \chi_o) \Big),
\ee
so even the states that are not initially excited with respect to the Fock vacuum contribute to the spatial volume in a similar way (i.e., with exponential growth) as the states that are initially in an excited condensate state.  Further, this shows that, for these $\vec \jmath, \vec m, \iota$, the contribution to the spatial volume undergoes a `bounce' with a minimal value of 0, but whose evolution always remains well-defined.

In the case that the initial state is $|\Psi(\chi_o) = |0\ket$ (i.e., the Fock vacuum for all $\vec\jmath, \vec m, \iota$), then the total spatial volume is simply given by a sum over all $\vec \jmath, \vec m, \iota$ of \eqref{vj0}, with the result being a bouncing universe with a minimum value of $V=0$.  In this case, however, the expectation value of the momentum of the scalar field $\chi$ is $\bra \pi_\chi \ket = 0$, so a state of this type does not have a simple classical interpretation since the energy density, which is classically given by $\ve = \pi_\chi^2 / 2V^2$, would be zero.

\bigskip

Comparing the dynamics of the expectation value of the volume $V$ and momentum of the massless scalar field $\pi_\chi$ to the classical solution $V = A \exp[\pm \sqrt{12 \pi G} (\chi-\chi_o)]$ (this can be derived from the relational form of the Friedmann equations given in, e.g., Appendix A1 of \cite{Oriti:2016qtz}) and $\pi_\chi$ constant, it is immediate to see that the solution \eqref{sol-v}, and also the solution \eqref{vj0}, has the correct classical limit on both sides of the bounce if all $M_{\vec \jmath, \vec m, \iota}^2 = 3 \pi G$, since for sufficiently large $|\chi|$ all of the contributions to $\bra V \ket_\sigma$ go as $\sim \exp(2 |M_{\vec \jmath, \vec m, \iota} \chi|)$.

However, it is not necessary that all $M_{\vec \jmath, \vec m, \iota}$ satisfy $M_{\vec \jmath, \vec m, \iota}^2 = 3 \pi G$ to obtain the correct classical limit.  If there is a maximal $|M_{\vec \jmath, \vec m, \iota}|$, then at sufficiently late relational times it will be these quanta that will dominate the dynamics and to recover the correct classical limit at late times it is sufficient that $M_{\vec \jmath, \vec m, \iota}^2 = 3 \pi G$ hold only for the largest $|M_{\vec \jmath, \vec m, \iota}|$.

Another possibility is that, depending on $\vec \jmath, \vec m, \iota$, in some cases $\mK^{(0)}$ and $\mK^{(2)}$ may have the same sign, while in other cases they may have opposite signs.  An example of this is for the coupling $\mK^{(0)}_{\vec \jmath, \vec m, \iota} = \alpha + \beta \sum_{i=1}^4 j_i (j_i+1)$ where one of $\alpha$ and $\beta$ is positive and the other is negative \cite{Gielen:2016uft}.  In general, in this case some of the contributions to the Hamiltonian will be of the form \eqref{ham-1} while others will be of the form \eqref{ham-2}.  For the first group, the Fock vacuum is stable (for a positive $M_{\vec \jmath, \vec m, \iota}$ and in the limit that the potential $U$ is negligible), but not for the second group.  For a `mixed' Hamiltonian of this type, consider a state $|\Psi\ket = |0\ket_1 \otimes |\sigma\ket_2$ where the first group is unexcited with respect to the Fock vacuum, but the second group is in a condensate state of the form \eqref{def-sigma}.  Then, the first family of quanta would never be excited throughout the evolution (except by the potential $U$ at subleading order), while the remainder would evolve as described earlier in this section.  In particular, so long as there is at least one GFT excitation with a contribution of the form \eqref{ham-2} to the Hamiltonian, the emergent space-time for the state $|0\ket_1 \otimes |\sigma\ket_2$ will undergo a non-singular bounce, with the correct classical limit on either side (assuming $M_{\vec \jmath, \vec m, \iota}^2 = 3 \pi G$ for that particular excitation).  This matches what has been found previously in a similar context \cite{Gielen:2016uft}.  (If the initial state is instead $|0\ket$, then so long as there is at least one contribution \eqref{ham-2} to the Hamiltonian, the emergent space-time will undergo a bounce through the no-space state $|0\ket$, but still with the correct limit either side of the bounce.)

This demonstrates that the cosmological sector imposes some constraints on the couplings in any relational GFT, but there exists a broad range of group field theories that can reproduce the correct classical limit in this cosmological sector.

\bigskip

It is also interesting to compare these results with those of loop quantum cosmology (LQC), where the background-independent, non-perturbative quantization techniques of loop quantum gravity are applied to the symmetry-reduced phase space of, e.g., FLRW space-times \cite{Ashtekar:2011ni, Banerjee:2011qu}.  One of the key steps in LQC is expressing the field strength operator in terms of a holonomy of the Ashtekar-Barbero connection around a loop of minimal area; the motivation underlying this is the assumption that a cosmological space-time is composed of a very large number of Planck-scale quanta of geometry, which can be approximated by $N \gg 1$ `minimally excited' quanta of geometry.  In LQC, for a sharply-peaked state the dynamics of the expectation value of the spatial volume $V$ for the flat FLRW space-time minimally coupled to a massless scalar field $\chi$, expressed in terms of the relational time $\chi$, is
\be
\bra V \ket_{LQC} = \f{\pi_\chi}{\sqrt{2 \ve_c}} \cosh \Big( \sqrt{12\pi G} (\chi-\chi_o) \Big),
\ee
where $\ve_c \sim 1/G^2 \hbar$ is the critical energy density the bounce occurs at.  (This can be derived from the relational effective equations of motion for LQC given in, e.g., Appendix A2 of \cite{Oriti:2016qtz}.)

In the GFT context considered here, the assumptions underlying LQC would correspond to a condensate state where a single type of excitation dominates the sum contributing to the total volume in \eqref{sol-v}.  For example, this would be the case if all of the contributions to the relational Hamiltonian are of the form \eqref{ham-1}, except for one particular configuration whose contribution the relational Hamiltonian is of the form \eqref{ham-2}; then this one configuration is the only unstable mode that will be excited by the relational dynamics.  If only one mode dominates the condensate state, then the sum in \eqref{sol-v} trivializes so that
\be
\bra V \ket_\sigma = \f{V_{\vec \jmath_o, \iota_o} \pi_\chi}{\hbar M_{\vec \jmath_o, \vec m_o, \iota_o}} \cosh \Big( 2M_{\vec \jmath_o, \vec m_o, \iota_o} (\chi-\chi_o) \Big),
\ee
where $A_{\vec \jmath_o, \vec m_o, \iota_o}$ has been rewritten using \eqref{exp-pi}.  Using $M_{\vec \jmath_o, \vec m_o, \iota_o}^2 = 3 \pi G$ to ensure the dynamics have the correct classical limit, these dynamics for $\bra V \ket_\sigma$ are identical to the effective LQC dynamics for $V$ with a critical energy density of
\be
\ve_c = \f{3 \pi G \hbar^2}{2 V_{\vec \jmath_o, \iota_o}^2}.
\ee
(Using the approximate scaling $V_{\vec \jmath, \iota} \sim \b\jmath^{\,3/2} \, \lp^3$ with $\b\jmath = (j_1 j_2 j_3 j_4)^{1/4}$, then $\ve_c \sim 3/(2 \b\jmath^3 G^2 \hbar)$.)  As a result, the bounce in the expectation value for the volume occurs when the energy density of the scalar field $\chi$ reaches the Planck scale, specifically $\ve_c$.

Note that in this case, it follows from \eqref{sol-v} and \eqref{exp-pi} (with only one term contributing in each of the sums) that the following effective Friedmann equation holds for the GFT condensate state:
\be
\left( \f{1}{3 \bra V \ket_\sigma} \f{d \bra V \ket_\sigma}{d\chi} \right)^2 = \f{4\pi G}{3} \left(1 - \f{1}{\ve_c} \f{\bra \pi_\chi\ket_\sigma^2}{2 \bra V \ket_\sigma^2} \right),
\ee
where the left-hand side is the square of the relational Hubble rate, and $\bra \ve \ket_\sigma = \bra \pi_\chi\ket_\sigma^2 / 2 \bra V \ket_\sigma^2$ corresponds to the energy density of the scalar field $\chi$.  This also agrees exactly with the effective Friedmann equation found in LQC expressed in terms of the relational clock $\chi$.  (While it is possible to calculate the Hubble rate for more general states, the resulting expression is not as simple since the sums in \eqref{sol-v} and \eqref{exp-pi} no longer trivialize.)

\bigskip

The results in this section are very similar to what has been found before in GFT, both in a framework that is not explicitly relational with a complex field $\vp$ with `timeless' commutation relations \cite{Oriti:2016qtz, Oriti:2016ueo}, and also in a toy model developed to capture the salient features required for a good cosmological sector \cite{Adjei:2017bfm}, namely `equal relational time' commutators and a squeezing operator as the relational Hamiltonian.  It is interesting that starting from the GFT action \eqref{gft-s}, it is possible to derive commutation relations and a Hamiltonian analogous to those postulated in \cite{Adjei:2017bfm}.  The results obtained here are slightly different from those found in \cite{Oriti:2016qtz, Oriti:2016ueo} using a different GFT based on a complex field and `timeless' commutation relations, with two main differences: (i) in \cite{Oriti:2016qtz, Oriti:2016ueo} there also appeared a new constant of the motion (which does not appear here) that if non-zero shifts the energy/curvature scale of the bounce, and (ii) here $\pi_\chi$ is captured by the relational energy, while in \cite{Oriti:2016qtz, Oriti:2016ueo} $\pi_\chi$ is related to the $U(1)$ symmetry of the complex GFT field.  Also, the solution for $\rho_{\vec\jmath, \vec m, \iota}(\chi)$ is simpler in the relational framework used here (compare \eqref{rho-sol} with results in \cite{deCesare:2016axk}).  Despite these differences, the qualitative results obtained for condensate states in these two different GFTs, namely a non-singular bouncing cosmology with the bounce occuring at the Planck scale, are similar and so suggest that these results are robust and independent of the details of the GFT.

\section{Discussion}
\label{s.disc}

For a group field theory corresponding to Lorentzian gravity minimally coupled to a massless scalar field $\chi$, a relational Hamiltonian for GFT can be obtained from the Legendre transform of the GFT Lagrangian using $\chi$ as the relational time variable.  This provides a new path to define the quantum theory for the GFT using `equal relational time' commutation rules for the fundamental operators corresponding to the GFT field $\vp$ and its conjugate momentum $\pi$.

As condensate states are by their nature homogeneous in that all quanta are in identical states, it has been conjectured that in GFT these states may correspond, when coarse-grained, to cosmological space-times.  In line with this expectation, from the dynamics of the simplest coherent states emerge the Friedmann equations for a flat FLRW space-time, with quantum gravity corrections that ensure that the big-bang singularity state is generically resolved and replaced by a bounce.  A sufficient condition for these Friedmann equations to have the correct classical limit is that the largest of the GFT couplings $M_{\vec \jmath, \vec m, \iota}$ satisfy $M_{\vec \jmath, \vec m, \iota}^2 = 3 \pi G$.

In addition, in the case that one particular excitation dominates the condensate state, then the dynamics are identical to those of loop quantum cosmology, with the critical energy density given by $\ve_c = 3 \pi G \hbar^2 / 2 V_{\vec \jmath_o, \iota_o}^2$, where $V_{\vec \jmath_o, \iota_o}$ denotes the volume of the dominant quanta.  This result further strengthens the results of loop quantum cosmology as qualitatively similar results (and even quantitatively identical results in a limiting case) are obtained in the full GFT quantum gravity theory that includes all degrees of freedom.

The predictions obtained for GFT condensate states are qualitatively very similar, whether the GFT is defined in a timeless framework with a complex field $\vp$, or in the explicitly relational framework considered here based on a real-valued field $\vp$, with a relational Hamiltonian and equal relational time commutation relations.  However, there are some advantages to working with the relational GFT, as calculations are typically simpler: in particular, the instability of the Fock vacuum $|0\ket$ for a Hamiltonian of the form \eqref{ham-2} is immediately clear, and it is also easier to work with operators evaluated at an instant of relational time $\chi$ in this setting (in the timeless framework, since $\chi$ is continuous, relational operators are distributional and must be smeared over some interval $\delta\chi$ to be well-defined).

It would be interesting to extend these results in a number of directions.  The simplest extension would be to consider a GFT where the kinetic term in the action is not diagonal in the group indices.  There has also been recent work in the non-relational form of GFT, as applied to cosmological condensates, to study the effect of the interaction term \cite{deCesare:2016rsf, Pithis:2016wzf}, to include anisotropies in the quanta of geometry \cite{deCesare:2017ynn}, and to handle cosmological perturbations \cite{Gielen:2017eco, Gerhardt:2018byq}; the same could be done in the relational framework developed here.

A potentially important generalization would be to determine whether it is possible to construct a relational Hamiltonian in the case that the scalar field has a non-vanishing potential, in which case the GFT action would depend explicitly on the relational time variable \cite{Li:2017uao}.  Another interesting extension would be to go beyond leading order in the expansion in $\hbar$ in \eqref{exp-k}.  However, this would give a higher-derivative theory which would likely prove difficult to quantize canonically.  Finally, it would be interesting to also consider the case where the GFT field $\vp$ is complex; this would imply the existence of `anti-particles' of geometry.  While the interpretation of these geometric `anti-particles' is not immediately obvious (although see \cite{Rovelli:2012yy, Christodoulou:2012sm, Oriti:2013aqa} for some ideas that might be relevant), it would nonetheless be interesting to understand how the presence of anti-particles may modify the theory and its predictions.

\bigskip

\noindent
{\it Acknowledgments:} 
%
%
I thank Marco de Cesare, Steffen Gielen, Viqar Husain, Daniele Oriti and Axel Polaczek for helpful discussions and comments on an earlier draft of the paper.
This work was supported in part by the Natural Science and Engineering Research Council of Canada.

\small
\raggedright

\end{document}